# PRISA: a simple software for determining refractive index, extinction co-efficient, dispersion energy, band gap, and thickness of semiconductor and dielectric thin films

S. Jena[*], R. B. Tokas, S. Thakur, D. V. Udupa

Atomic & Molecular Physics Division, Bhabha Atomic Research Centre, Mumbai 400 085, India
*shuvendujena9@gmail.com, shujena@barc.gov.in

**Abstract:** A simple user-friendly software named PRISA has been developed to determine optical constants (refractive index and extinction co-efficient), dispersion parameters (oscillator energy and dispersion energy), absorption co-efficient, band gap and thickness of semiconductor and dielectric thin films from their measured transmission spectrum, only. The thickness, refractive index, and extinction co-efficient of the films have been derived using Envelope method proposed by Swanepoel. The absorption co-efficient in the strong absorption region is calculated using the method proposed by Connel and Lewis. Subsequently, both direct and indirect bandgap of the films is estimated from the absorption co-efficient spectrum using Tauc plot. The codes for the software are written in Python and the graphical user interface is programmed with tkinter package of Python. It provides convenient input and output of the measured and derived data. The software has a feature to retrieve transmission spectrum using the derived parameters in order to check their reliability. The performance of the software is verified by analyzing numerically generated transmission spectra of a-Si:H amorphous semiconductor thin films, and experimentally measured transmission spectra of electron beam evaporated $HfO_2$ dielectric thin films as examples. PRISA is found to be much simpler and accurate as compared to the other freely available softwares. To help other researchers working on thin films, the software is made freely available at https://www.shuvendujena.tk/download.

## 1. Introduction

Performance of most thin film optical coatings and optoelectronic devices depends on optical properties and thickness of thin films [1]. For example, dielectric thin films such as $HfO_2$, $TiO_2$, $ZrO_2$, $Ta_2O_5$, $SiO_2$, etc. [2] are used for developing thin film multilayer coating devices such as antireflection coatings, high reflection mirrors, bandpass filters, beam splitters, beam combiners, and heat mirrors [3]. The optical properties such as refractive index, extinction co-efficient, and thickness of the oxide thin films are very crucial for the design and development of such optical coating devices [4]. Optical properties of amorphous semiconductor materials such as a-Si, a-Si:H, a-Ge:H, $As_2Se_3$, $GeS_2$, $Sb_2Se_3$, etc. are exploited for numerous applications over the years. For example, hydrogenated amorphous silicon (a-Si:H) thin film is widely used in solar cells, while thin film of chalcogenide glass like $As_2Se_3$ is used in optical memory devices [5]. The high sensitivity of optical bandgap and refractive index of the semiconductor chalcogenide thin films to external factors such as laser, neutron, electron, and gamma irradiation has gathered significant attention over few decades [6]. These photosensitive natures of such amorphous semiconductor thin films have found numerous applications in optoelectronics, optical imaging and recording, information storage, absorption filters, high-resolution display devices, acousto-optic devices, and photolithography [7]. Therefore, precise knowledge of optical properties of such thin films is very much essential for ever demanding technology of advanced optics and photonics.

In recent years, numerous methods such as spectroscopic ellipsometry [8], inverse synthesis or fitting method [9-11], reflection spectrum method [12-14], etc. have been developed for determining optical properties and thickness of thin films. However, the Envelope method proposed by Swanepoel [15] still remains the simplest approach for deriving refractive index ($n$), extinction co-efficient ($k$), and thickness ($d$) of thin films from the measured transmission spectrum. Therefore, Envelope/Swanepoel method has been widely used by several researchers for optical characterization of thin films. Different analytical and numerical algorithms based on Envelope method have been proposed and demonstrated by some researchers for better analysis of thin films [16-18]. We have implemented the optimized analytical algorithms to get the most accurate value of thickness and refractive index of thin films, which is demonstrated in later part of this manuscript. The other important optical parameters that can be derived from refractive index spectra are dispersion energy ($E_d$), and oscillator energy ($E_0$) using single oscillator model, which provides information about the structural and chemical properties of the materials [19-21]. Wemple and DiDomenico [20] have shown that dispersive refractive index of many materials can be precisely defined by just varying the strength and position of a single oscillator. This is the consequence of a simple empirical relation obeyed by $E_d$, which shows that iconicity, coordination number, and chemical valence play key roles in defining the behavior of refractive indices, and, subsequently, electro-optic, nonlinear-optical, and photo-elastic effects [20]. The parameter $E_d$ controls the interaction potential describing these optical effects and, thus, directly influences their magnitudes. The parameter $E_0$ is related to the bandgap of materials. Optical bandgap value ($E_g$) of thin film materials is a very crucial parameter that decides whether the thin film could be useful for solar cell, optical filter, and opto-electronic devices applications or not. Therefore, exact knowledge of bandgap (direct/indirect) of thin films are very much essential, and is derived from the transmission edge of the spectrum using Tauc plots.

Transmission spectrum is analyzed to derive the optical parameters and thickness of thin films. The analysis/calculation requires data processing, and managing of different files and formats. Researchers use softwares like Mathematica, MATLAB, Mathcad, Maple, Origin, Excel, etc. for such calculations. Among them, some are tedious, and some require expensive licenses or programming knowledge. As a result, it takes a lot of time in script creating process, programming, and then managing files taken from different spectrophotometer instruments. There are several commercially available softwares/programs for transmission analysis such as TFCalc [22], FilmStar [23], OptiLayer [24], Essential Macleod [25], Film Wizard [26], and others, but they are expensive. There exist several freeware or open source softwares such as OpenFilter [27], RefFIT [28], FreeSnell [29], Optical [30], and PUMA [31], but they are not much user friendly. Most of the softwares use inverse synthesis method to derive optical parameters from measured spectrum. Inverse synthesis involves different dispersion models and nonlinear optimization programming, and required prior guess of film thickness, whereas Envelope method is very simple, and does not require any such information. One software named PARAV [32] based on Envelope method is freely available, but gives inaccurate result, which is verified in the present manuscript. Moreover, all the above mentioned commercial and free software do not have the features to estimate dispersion energy parameters ($E_d$, $E_0$, and $n_0$), Tauc plots, and band gap ($E_g$) of thin films. This encouraged us to develop a simple and user-friendly graphical user interface (GUI) software named PRISA that can derive $n$, $k$, $E_d$, $E_0$, $α$, $E_g$, and $d$ of thin films from the measured transmission spectrum, which will be useful to the researchers working in the area of thin films. The software is developed using Python [33], and the GUI is implemented with tkinter package [34]. The reliability and performance of the software is tested by analyzing a semiconductor thin film (a-Si:H), and a dielectric thin film ($HfO_2$) as examples.

The paper is organized as follows: Sec. 2 provides the theoretical formulations used for the algorithm of the software. Sec. 3 clearly describes the complete GUI of the software.

Sec. 4 verifies the reliability and performance of the software. Details of the software distribution, and summary of the work are presented in Sec.5, and Sec. 6, respectively.

## 2. Theoretical background

### 2.1. Optical constants and thickness

Fig. 1(a) shows the thin film/substrate structure assigned with corresponding optical parameters and thickness. The substrate is transparent and the film can be semitransparent or transparent. The $n$, $k$, and $d$ are refractive index, extinction co-efficient and thickness of the thin film, while $n_s$, $k_s=0$ and $d_s$ are that of the substrate. The refractive index of air is assumed to be $n_0=1$. If the film thickness is uniform, then the interference effects give rise to a transmission spectrum as shown in Fig. 1(b). These interference fringes can be used to determine the optical constants and thickness of thin films using Envelope method, which has been developed and refined by several workers after the work of Manifacier *et al.* [35]. This method uses the interference maxima and minima to derive $n$, $k$ and $d$ of the film.

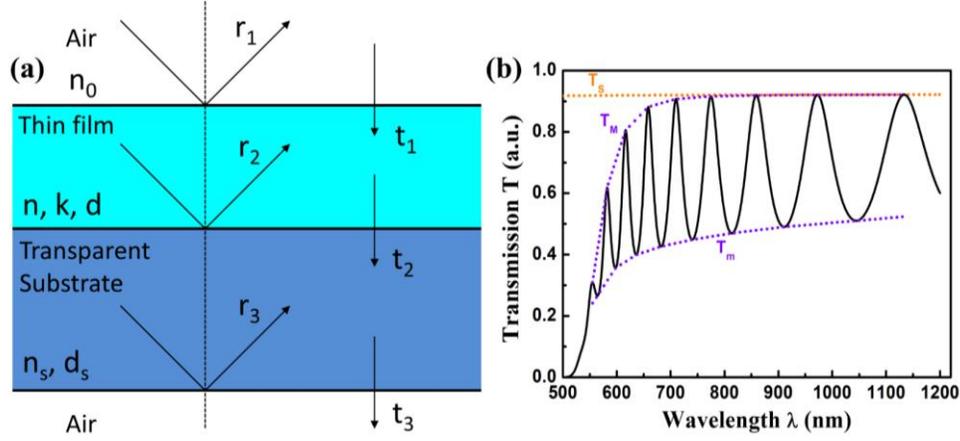

Fig. 1. (a) Schematic of thin film/substrate system with relevant optical parameters and thicknesses used in the calculation of Envelope method and retrieving transmission spectrum. (b) Transmission spectrum of a-Si:H thin film on glass substrate. Curves $T_S$, $T_M$, and $T_m$, are as per the text.

The extremes in a transmission spectrum occur according to the interference condition

$$2nd = m\lambda \tag{1}$$

where $m$ is an integer for maxima and half integer for minima, and $\lambda$ is the corresponding wavelength. The curve $T_M$ and $T_m$ define the envelopes passing tangentially through the maxima and minima in the spectrum as shown in Fig. 1(b). Generally, transmission spectrum has two regions as per absorption in the film: region of weak and medium absorption, and strong absorption region. The first approximation of refractive index of thin film at wavelength of different extreme points is given by

$$n = \sqrt{N + \sqrt{N^2 - n_s^2}} \tag{2}$$

where $\quad N = 2n_s \dfrac{T_M - T_m}{T_M T_m} + \dfrac{n_s^2 + 1}{2}$ (3)

If $n_1$ and $n_2$ are the refractive indices at two adjacent maxima (or minima) at wavelengths $\lambda_1$ and $\lambda_2$ with $\lambda_1 > \lambda_2$, then the film thickness using equation (1) can be estimated by the relation:

$$d = \frac{\lambda_1 \lambda_2}{2[\lambda_1 n(\lambda_1) - \lambda_2 n(\lambda_2)]} \tag{4}$$

Exact smooth envelopes of $T_M$ and $T_m$ are always difficult to generate and may have little errors. As a result, the initial estimated values of refractive index ($n_{in}$) and thickness ($d_{in}$) of the films derived from equations (2) and (4) could be inaccurate. In order to get accurate values of refractive index and thickness of the thin film, following optimized algorithm should be executed. At first, use the average value of $d_{in}$ obtained from each two adjacent extremes in equation (1) to calculate order number ($m_{in}$) for each extreme and round off the obtained $m_{in}$ to the nearby integer or half integer. These rounded values are accepted as the exact order number $m_{fn}$ corresponding to each maxima or minima. Then, the values of $m_{fn}$ and $n_{in}$ are used in equation (1) to estimate the accurate thickness $d_{fn}$ for each extreme excluding first maximum and minimum generally present in the strong absorption region. The average value of $d_{fn}$ is finally accepted as true thickness of the film. Finally, the true thickness and the exact $m$ values are used in equation (1) to derive the accurate refractive index $n_{fn}$ for each extreme. Once $n(\lambda)$ and $d$ of the film are known, then the extinction co-efficient ($k$) as a function of wavelength $\lambda$ can be calculated as follows:

$$k = -\frac{\lambda}{4\pi d} \ln(x) \tag{5}$$

where $x = \dfrac{F - \sqrt{F^2 - [(n^2-1)^3(n^2-n_s^4)]}}{(n-1)^3(n-n_s^2)}$ (5a)

and $F = 4n^2 n_s \dfrac{T_M + T_m}{T_M T_m}$ (5b)

The refractive index and extinction co-efficient as a function of wavelength i.e. $n(\lambda)$ and $k(\lambda)$ can be extrapolated by fitting the estimated data using suitable dispersion models such as Cauchy or Sellimier dispersion equations [36]. It is noteworthy to mention that Envelope method works fine and gives very accurate results for relatively thick films as they exhibit large number of interference fringes in their transmission spectrum, while it does not work for metallic or highly absorbing thin films due to absence of interference fringes. It is also not effective for thin films having local absorption bands in between interference fringes. The accuracy of the method becomes poor for lower thickness films. Because the number of fringe extremes gets lower and the spacing between the fringe extremes gets wider with decreasing thickness, which makes interpolation between extremes more difficult [37]. Instead of these shortcoming, Envelope method still remains the popular as a routine analysis technique for estimation of optical constants of transparent or semiconducting thin/thick films.

2.2.  Retrieval of transmission spectrum

The transmission ($T$) of a single layer homogeneous thin film deposited on a thick transparent substrate, whose schematic is shown in Fig. 1(a), is given by [38]:

$$T = \frac{n_s |E_t|^2 (1-r_3^2)}{n_0 (1-|r_2 r_3|^2)} \tag{6}$$

where $E_t = \dfrac{t_1 t_2 e^{-i\beta}}{1 + r_1 r_2 e^{-2i\beta}}$ (6a)

$$r_1 = \frac{n_0 - \tilde{n}}{n_0 + \tilde{n}} \qquad r_2 = \frac{\tilde{n} - n_s}{\tilde{n} + n_s} \qquad r_3 = \frac{n_s - n_0}{n_s + n_0} \tag{6b}$$

$$t_1 = \frac{2n_0}{n_0 + \tilde{n}} \qquad t_2 = \frac{2\tilde{n}}{\tilde{n} + n_s} \qquad t_3 = \frac{2n_s}{n_s + n_0} \tag{6c}$$

$$\beta = \frac{2\pi \tilde{n} d}{\lambda} \tag{6d}$$

Here $\tilde{n} = n - ik$ is the complex refractive index of thin film. $n$, $k$ and $d$ are refractive index, extinction co-efficient, and thickness of the thin film. $(r_1, t_1)$, $(r_2, t_2)$, and $(r_3, t_3)$ are the amplitudes of (reflection, transmission) co-efficients of air-film, film-substrate, and substrate-air interfaces, respectively. Equation (6) can be used to check the reliability of the $n$, $k$, and $d$ values obtained using Envelope/Swanepoel method. These optical parameters and thickness values will be used to simulate or retrieve the transmission spectrum. If the theoretically generated/retrieved spectrum overlaps the original measured spectrum, then it confirms the reliability of the obtained $n$, $k$, and $d$. If it does not then either the thin film is inhomogeneous for which the adopted Swanepoel method does not work or the given refractive index for the substrate is wrong or it could be both.

## 2.3. Dispersion energy parameters

Wemple and DiDomenico have shown that single oscillator model can successfully describe the dielectric response function of various materials. Using time-dependent perturbation theory and considering only a single group of valence and conduction bands, the frequency dependent real part of the dielectric permittivity of a material can be written as [20]

$$\varepsilon_1(\omega) = 1 + \frac{4\pi e^2}{m\Omega} \sum_{c,v} \frac{f_{cv}^\alpha(\vec{k})}{\omega_{cv}^2(\vec{k}) - \omega^2} \tag{7}$$

where, $e$ and $m$ are charge and mass of electron, respectively. $\Omega$ is the volume of the crystal $f_{cv}^\alpha(\vec{k})$ is the interband oscillator strength, and the sum extends over only valence and conduction bands. If we consider the dispersive dielectric permittivity is due to interband transitions that are described by individual oscillators, and each valence electron contributes one such oscillator, then equation (7) can be approximately given by

$$\varepsilon_1(\omega) = 1 + \omega_p^2 \sum_s \frac{f_s}{\omega_s^2 - \omega^2} \tag{8}$$

Here, $\omega_p$ is the plasma frequency, $f_s$ is the electric-dipole oscillator strength associated with transitions at frequency $\omega_s$. For $\omega < \omega_s$, the above equation can be written as

$$\varepsilon_1(\omega) = 1 + \omega_p^2 \left[ \frac{f_1}{\omega_1^2 - \omega^2} + \sum_{s \neq 1} \frac{f_s}{\omega_s^2} \left(1 + \frac{\omega^2}{\omega_s^2}\right) \right] \tag{9}$$

Equation (9) can be approximated to a single oscillator by adding all the higher order contributions with the 1$^{st}$ resonant oscillator ($s=1$) and retaining terms to order $\omega^2$, which yields the following equation

$$\varepsilon_1(\omega) \approx 1 + \frac{F}{E_0^2 - (\hbar\omega)^2} \tag{10}$$

where the parameters $E_0$, and $F$ are directly related to all the $f_s$ and $\omega_s$ in equation (9). Wemple and DiDomenico have experimentally shown that the above two parameters are related to each other through a simple equation $F=E_0E_d$, where $E_d$ is the so-called dispersion energy, and $E_0$ is the single oscillator energy. Finally, equation (10) can be expressed as

$$\varepsilon_1(E) = n^2(E) = 1 + \frac{E_d E_0}{E_0^2 - E^2} \tag{11}$$

$$\frac{1}{n^2(E) - 1} = \left(\frac{-1}{E_d E_0}\right) E^2 + \left(\frac{E_0}{E_d}\right) \tag{12}$$

where $E=\hbar\omega$ (eV) is the energy of the light. By plotting $(n^2-1)^{-1}$ versus $E^2$ and fitting it with a straight line, the value of the two meaningful parameters $E_d$ and $E_0$ can be obtained from the slope $-1/E_dE_0$ and the intercept $E_0/E_d$, whose physical significances are discussed as follows.

The dispersion energy $E_d$ is a measure of the strength of interband optical transitions, and is related to the charge distribution within unit cell and hence with the chemical bonding. It is directly related to physical parameters of the material through the empirical relation $E_d=\beta N_c Z_a N_e$, where $N_c$ is the effective co-ordination number of the cation nearest neighbor to the anion, $Z_a$ is the formal chemical valence of the anion, $N_e$ is the effective number of valence electrons per anion (cores excluded), and $\beta$ is an energy constant expressing the covalent or ionic character of the solid through the values 0.37±0.04 and 0.26±0.03 eV, respectively [39]. The empirical relation indicates that the dispersion energy parameter $E_d$ is strongly related to the atomic/electronic structure environment in the thin film. Therefore, any variation in structure (chemical bonding, lattice structure, etc.) of thin films can be reflected in the value of $E_d$. The single oscillator energy $E_0$ is generally considered as an "average" energy-gap parameter equivalent to the energy difference between the "centres of gravity" of the valence and conduction bands. This is like the energy parameter ("Penn gap"), used by Penn' for the determination of the static refractive index ($n_0$) in semiconductors [3]. The parameter $E_0$ is empirically related to the lowest direct band gap ($E_g$) of the material through the equation $E_0 \approx 1.5E_g$ [40]. The static dielectric constant of any material is defined as $\varepsilon_0 = \lim_{E \to 0} n^2(E) = n_0^2$. The static refractive index ($n_0$) can be expressed in terms of dispersion parameters using equation (11) as $\varepsilon_0 = n_0^2 = 1 + (E_d/E_0)$. So, the knowledge of dispersion parameters allows us to determine the static dielectric constant of materials.

### 2.4. Absorption co-efficient and optical band gap

Absorption co-efficient ($\alpha$) can be calculated from the interference-free transmission spectrum $T_\alpha$ using the well-known method suggested by Connel and Lewis [41]. The simplified equations to determine $\alpha$ given in the reference [42] and appendix of the reference [15] are as follows:

$$\alpha = -\frac{1}{d}\ln\left(\frac{P + \sqrt{P^2 + 2QT_\alpha(1 - R_2R_3)}}{Q}\right) \tag{13}$$

where $P = (R_1 - 1)(R_2 - 1)(R_3 - 1)$ \hfill (13a)

$$Q = 2T_\alpha (R_1 R_2 + R_3 R_1 - 2R_1 R_2 R_3) \qquad (13b)$$

$$R_1 = r_1^2 \qquad R_2 = r_2^2 \qquad R_3 = r_3^2 \qquad (13c)$$

Where $r_1$, $r_2$, and $r_3$, are the Fresnel reflection co-efficients of the air-film, film-substrate, substrate-air interfaces, respectively as illustrated in equation (6b) earlier. Since, the value of imaginary part of the complex refractive index *i.e.* extinction co-efficient ($k$) is much less than its real part ($n$) in the region for which the value of absorption co-efficient $\alpha \leq 10^5 \text{cm}^{-1}$, therefore the value of $k$ is neglected in the expression of reflectivity at all interfaces. The geometric mean of transmission spectrum having interference fringes in the transparent region is $T_\alpha = (T_M T_m)^{1/2}$. But in the strong absorption region, the fringes vanish and all the three curves $T_\alpha$, $T_M$, and $T_m$, converge to a single curve.

The optical band gap ($E_g$) can be derived from the absorption co-efficient spectrum ($\alpha$) in the high absorption region ($\alpha \geq 10^{-4} \text{cm}^{-1}$) using the methods proposed by Tauc *et al.* [43], which was further developed by Mott and Davis [44]. They reported that the optical absorption strength is proportional to the difference between photon energy and band gap as follows:

$$(\alpha E)^{1/M} = A(E - E_g) \qquad (14)$$

where $E = h\nu$ is the photon's energy ($h$ is Planck's constant, and $\nu$ is the photon's frequency), $E_g$ is the band gap, and A is a proportionality constant. The value of the exponent denotes the nature of the electronic transition. The value of $M = 1/2$ and 2 represents direct and indirect allowed transitions, respectively while $M = 3/2$ and 3 represents direct and indirect forbidden transitions, respectively [45]. Usually, the allowed transitions dictate the fundamental absorption processes, giving either $M = 1/2$ or $M = 2$, for direct and indirect transitions, respectively. Thus, the basic procedure for a Tauc analysis is to acquire optical absorbance data for the sample in question that spans a range of energies from below the band-gap transition to above it. Plotting the $(\alpha E)^{1/2}$ versus $E$ and $(\alpha E)^2$ versus $E$ is a matter of testing $M = 1/2$ or $M = 2$ to compare which provides the better linear fit and thus identifies the correct transition type. Generally, absorption becomes saturate at higher energy and the curve deviates from linear in case of both direct and indirect bandgap thin films. The linear region for extrapolation should be wisely selected considering the reasons for the deviations in the lower and higher energy regions. Defect states near the band edge are responsible for the deviation in the low energy end, while saturation of transition states is responsible in higher energy end. The tail or edge associated with the defect states are popularly known as "Urbach tail" [46] and its distribution of density of sates is represented by an exponential function which can be evident from absorption curve of numerous thin films.

### 3. Software interface

PRISA software is written in Python 3.7, and uses scipy, numpy, and matplotlib modules in its coding. The graphical user interface (GUI) of the software is developed using tkinter package, and it appears as Fig. 2. The graphics of the interface may slightly change with respect to different computers. Fig. 2 is a screen short of the software in a DELL Inspiron laptop having Windows 7 operating system. In this section, we will outline the operation flow of the GUI of the software for analysing measured transmission spectrum.

At first, the spectrophotometric data to be imported must be .txt or .ascii formatted file with two column data of wavelength and transmission separated by a tab or space. The software automatically ignores the initial two row headers of the file while importing the data. The wavelength unit could be either nanometer (nm) or micron (µm), while the transmission unit could be either percentage (%) or absolute unit (a.u.). After choosing the unit of x-axis (wavelength) and y-axis (transmission), the data is imported on clicking "*Import data*" button in the GUI. Once the transmission spectrum data is imported, then the fringe maxima and minima of the spectrum can be selected, and their corresponding upper and lower envelope are generated, automatically by clicking the "*Make envelope*" button of the GUI. The envelopes are made using cubic-spline interpolation. For our calculation, the refractive index of the incident medium (air) is $n_0 \approx 1$. The exit medium is a transparent substrate. Generally, the value of substrate refractive index (*Sub. index ns*) is assumed as $n_s \approx 1.51$ for glass, and 1.47 for fused silica, respectively in the visible region. Now, the envelope data at extreme points together with incident and exit medium refractive indices are used to estimate thickness, refractive index ($n$), and extinction co-efficient ($k$) of the thin film in the weak and medium absorption region of the spectrum by clicking on "*Determine (n, k, & d)*" button. The $n$ and $k$ spectra are fitted and extrapolated to wider spectral range using following equations: $n(\lambda) = a_1 + (a_2/\lambda^2) + (a_3/\lambda^4)$, and $k(\lambda) = b_1 + (b_2/\lambda^2) + (b_3/\lambda^4)$. The obtained refractive index spectrum *i.e.* $n(\lambda)$ is used to derive dispersion parameters $E_d$ and $E_0$, and static refractive index $n_0$ using equation (12) as discussed above by clicking the "*Determine $E_d$, & $E_0$*" button of the GUI.

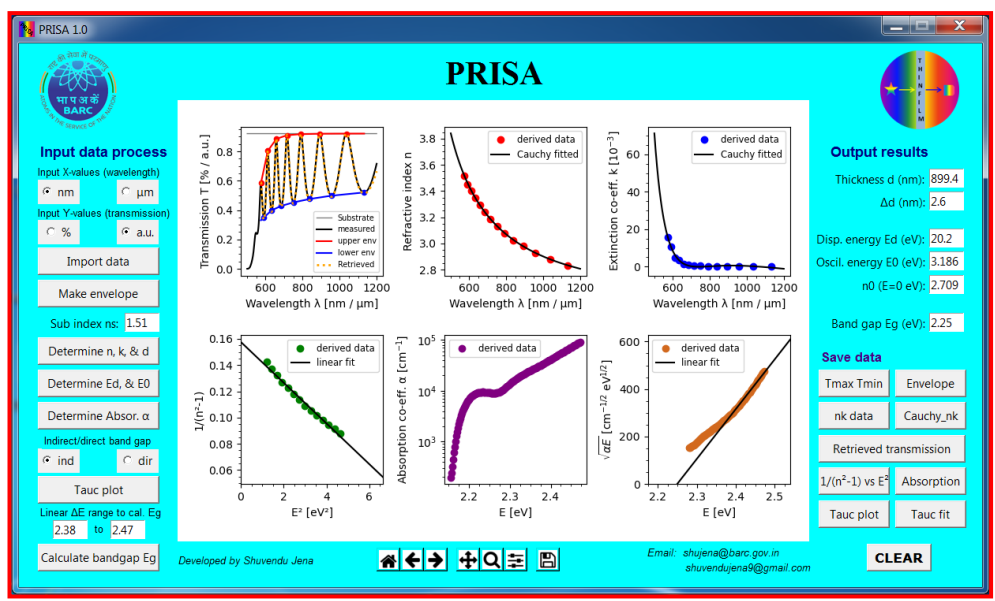

Fig. 2. GUI of PRISA software illustrating complete analysis of 900 nm thick a-Si:H film.

Now, absorption co-efficient ($\alpha$) in the strong absorption region is determined from the measured transmission spectrum using Lewis and Connel method as mentioned earlier by clicking the "*Determine Absor. α*" button. Subsequently, Tauc plot is drawn choosing either direct (*dir.* radio button) or indirect (*ind.* radio button) band gap by clicking the "*Tauc plot*" button of the GUI. The linear range in the Tauc plot for direct or indirect band gap can be carefully chosen by zooming out the Tauc plot, and its initial and final values are placed in the boxes under "*Linear ΔE range to cal. $E_g$*". Now, the corresponding linear range in the Tauc plot is fitted with equation (14) to determine the optical bandgap of the thin film by clicking the "*calculate bandgap $E_g$*" button in the software. All the plots can be scaled as well as zoomed in and out in the GUI itself. Different output results can be saved as *text* or *ascii* files by clicking the corresponding result buttons under "*Save data*" command of the GUI. The output files can be directly read by other plotting software like origin, excel, etc.

## 4. Verification and performance

Performance and reliability of PRISA is verified by analyzing numerically generated transmission spectra of a-Si:H thin films, and experimentally measured spectrum of electron beam (EB) evaporated $HfO_2$ thin films. The software has options to retrieve the transmission spectrum using $n$, $k$, & $d$ values obtained from PRISA using the expressions for air/thin film/substrate system. If the retrieved data exactly overlap the original spectrum, then the results can be evidently reliable. Subsequently, the results obtained from PRISA, and PARAV [47] softwares are compared.

### 4.1. Numerical spectrum of a-Si:H semiconductor thin film

The transmission spectrum of a-Si:H thin film is numerically generated using the optical properties as mentioned in the references [15, 48]. The $n$, and $k$ of the film are expressed by the following equations:

Refractive index: $n = \dfrac{3 \times 10^5}{\lambda^2} + 2.6$ (15)

Absorption co-efficient: $\log_{10} \alpha = \dfrac{1.5 \times 10^6}{\lambda^2} - 8$ (16)

Extinction co-efficient: $k = \dfrac{\alpha \lambda}{4\pi}$ (17)

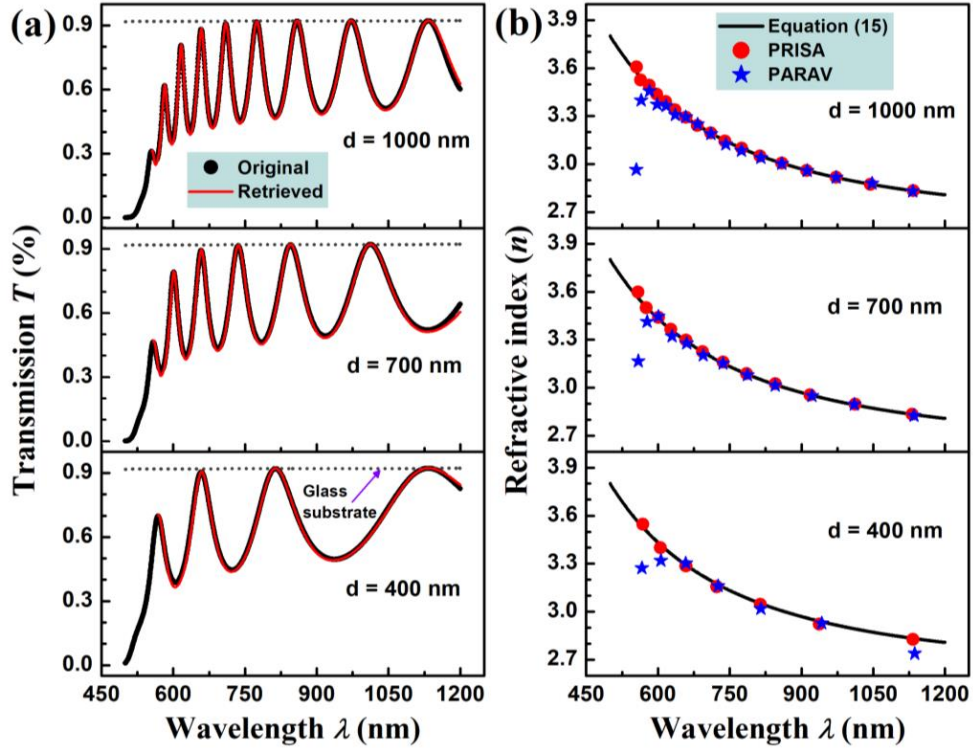

**Fig. 3.** (a) Transmission spectrum of a-Si:H thin film with different thicknesses $d$ =1000 nm, 700 nm, and 400 nm. The solid circle (●) represents the numerically generated spectrum, sold red line (——) represents the spectrum retrieved using the $n$, $k$, & $d$ values obtained from PRISA, and dotted grey line (······) represents substrate spectrum. (b) Refractive index spectrum of a-Si:H thin film derived from the corresponding transmission spectrum for different thicknesses using PRISA, and PARAV softwares. Equation (15) gives the actual refractive index (——) of a-Si:H, which is plotted to compare the accuracy of the two softwares.

The above expression for $n$ and $k$ matches to the experimentally measured values for a-Si:H [42]. Equation (6) is used to generate numerical transmission spectra of a-Si:H thin films of three different thicknesses d = 1000 nm, 700 nm, and 300 nm as shown in Fig. 3(a). Now, these transmission spectra are analyzed using PRISA. The refractive index spectra obtained from the analysis are plotted in Fig. 3(b) along with the $n(\lambda)$ data of equation (15), and that obtained using PARAV software. This figure clearly shows that PRISA gives much accurate results as compared to PARAV, and almost overlaps with the original $n(\lambda)$ data. The $n$ and $k$ values obtained using Envelope method are extrapolated using the dispersion relations mentioned earlier in section 3. These extrapolated $n$ & $k$, and $d$ are used in in equation (6) to retrieve the transmission spectrum in the medium-weak absorption and transparent region. The retrieved spectra for a-Si:H thin films for different thicknesses exactly overlap the original spectra as shown in Fig. 3(a), which confirms the reliability of the obtained value of $n$, $k$, and $d$ using PRISA. Accuracy of Envelope method primarily depends on the number of fringes in the transmission spectrum. More the number of fringes, better is the accuracy. Therefore, the software is tested for a-Si:H thin films from lower to higher number of interference fringes. Higher the film thickness, larger is the number of fringes in the spectrum. PRISA is implemented to analyses numerically generated transmission spectra of a-Si:H thin films with film thicknesses varying from 300-1500 nm. The obtained $n$ and $k$ values are plotted in Fig. (4) along with their numerical data obtained from equation (15) and (17), respectively. The figure shows that PRISA performs well for smaller to larger thickness films. The extinction co-efficient $k(\lambda)$ spectra does not match to the numerical data in the transparent region, which shows that the reliability of $k$ value becomes poor above ~$10^{-4}$, which is one of the limitations of Envelope method. Now, the dispersive refractive index spectra for different thicknesses of a-Si:H thin films are analyzed using equation (12) to obtain their corresponding $E_d$, $E_0$, and $n_0$. The derived dispersion energy parameters with varying thicknesses are plotted in Fig. 4 (c). It can be clearly seen that the values obtained for higher thickness films are closer to the actual values of the dispersion parameters, while the values are slightly deviated for the lower thickness values. This is primarily due to lower number of data points in the $1/n^2-1$ vs. $E$ plot for lower thickness films, consequently the linear fitting to that data with the equation (13) gets poor. As a result, the parameters $E_d$, and $E_0$ derived from the fitting parameters of slope and intercept becomes deviated from the actual value for lower thickness films.

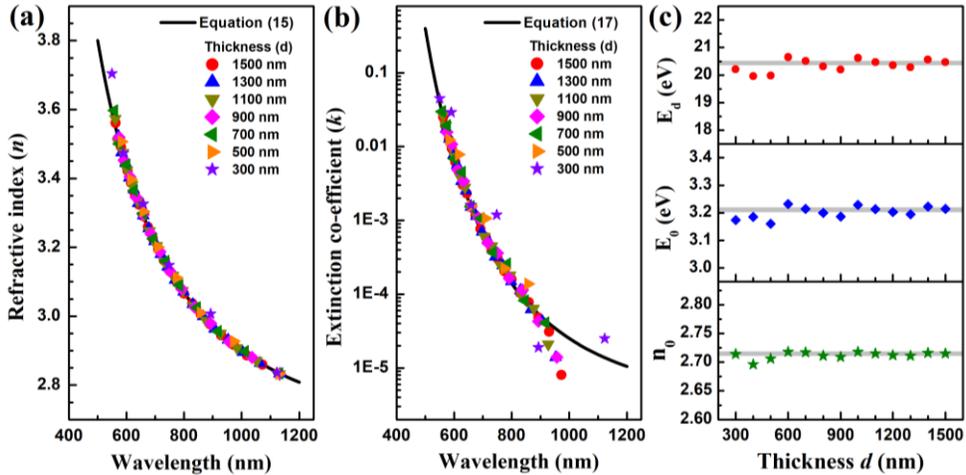

Fig. 4. (a) Refractive index ($n$) and (b) extinction co-efficient ($k$) spectra of a-Si:H thin film for different film thicknesses ranging from 300 to 1500 nm. Symbols (●, ▲, ▼, ◄, ►, etc.) are the spectra derived using PRISA, and solid black line (——) spectrum for $n$, and $k$ are calculated using equation (21), and (23), respectively. (c) Dispersion energy ($E_d$), oscillator energy ($E_0$), and static refractive index ($n_0$) of a-Si:H thin film with thickness varying from 300 to 1500 nm obtained using PRISA. The grey strip (——) represents their corresponding true values.

The thickness of a-Si:H thin films obtained from PRISA and PARAV softwares are listed in Table-1 along with the thickness data used to generate the numerical spectrum. It is clearly evident that PRISA gives more accurate thickness result as compared to that of PARAV for a given thin film sample. The thickness value derived using PRISA closely matches to the true thickness value within error bar, while PARAV gives thickness values with large deviation from the original value. Since the refractive index value obtained from PARAV largely deviates in the lower wavelength region of the spectrum as shown in Fig. 3(b), which subsequently affects the thickness values as per equation (1). The algorithm in PRISA avoids 1$^{st}$ maximum and 1$^{st}$ minimum in the strong absorption region to get accurate thickness value of the film. As per equation (4), at least two peak maxima are required for thickness estimation. Therefore, minimum three numbers of peaks in transmission spectrum are necessary for the optical analysis using PRISA. For a-Si:H thin films thickness below 200 nm, the number of peaks in the transmission spectrum goes below three, therefore the software fails to estimate the optical constants of the films. Hence, the analysis is done up to a-Si:H thin films of thickness 300 nm.

**Table-1: Comparison of thickness values of a-Si:H thin film obtained using PRISA and PARAV softwares for different film thicknesses ranging from 1500 to 300 nm.**

| a-Si:H thin film sample name | Thickness (nm) used to generate spectrum | Derived thickness (nm) | |
|---|---|---|---|
| | | PRISA | PARAV |
| S-1 | 1500 | 1500.5 ± 5.4 | 1615 |
| S-2 | 1400 | 1400.7 ± 6 | 1546 |
| S-3 | 1300 | 1300.1 ± 3.8 | 1379 |
| S-4 | 1200 | 1200.4 ± 4 | 1334 |
| S-5 | 1100 | 1099.9 ± 4.7 | 1214 |
| S-6 | 1000 | 999.9 ± 5.8 | 1164 |
| S-7 | 900 | 899.4 ± 2.6 | 1024 |
| S-8 | 800 | 799.9 ± 3.3 | 958 |
| S-9 | 700 | 698.8 ± 3.8 | 816 |
| S-10 | 600 | 599.4 ± 5.6 | 759 |
| S-11 | 500 | 498.3 ± 1.2 | 619 |
| S-12 | 400 | 400.8 ± 5.4 | 559 |
| S-13 | 300 | 297 ± 1.3 | 432 |

### 4.2. Experimental spectrum of HfO$_2$ dielectric thin film

A set of HfO$_2$ thin films were prepared using reactive electron beam evaporation at different O$_2$ pressure, and their optical transmission spectra were measured using UV-VIS-NIR spectrophotometer, whose details can be found in the reference [2]. A representative transmission spectrum of HfO$_2$ thin films (sample H-3) along with that of fused silica substrate is plotted in Fig. 5(a). Transmission spectra of all the films have been analyzed using PRISA (see supplementary material for complete analysis of the representative sample H-3, Fig. S1). Thickness of the HfO$_2$ thin films obtained using PRISA, and PARAV are listed in Table-2 along with the actual values which is estimated using inverse synthesis of transmission spectrum as mentioned in the reference [2]. The table shows that PRISA gives thickness values close to the actual values within error bar as compared to PARAV. The refractive index spectra of the sample H-3 are plotted in Fig. 5(b), which shows that PRISA gives more accurate refractive index spectrum as compared to PARAV. In the lower wavelength region, PARAV fails to match the true refractive index value. The $n(\lambda)$ curve obtained using PRISA shows slight deviation from the true values, which is primarily due to assumed refractive index value of the fused silica substrate. The HfO$_2$ thin film is deposited on fused silica substrate. Since the fused silica has refractive index variation of 1.54 to 1.46 in the wavelength region of 200-1200 nm, therefore, the assumption of refractive index value of 1.47 for the substrate gives slightly deviated values of refractive index values and thickness. Envelope method provides accurate results for substrate having less dispersive refractive index which means transmission of the substrate is almost parallel to the wavelength axis.

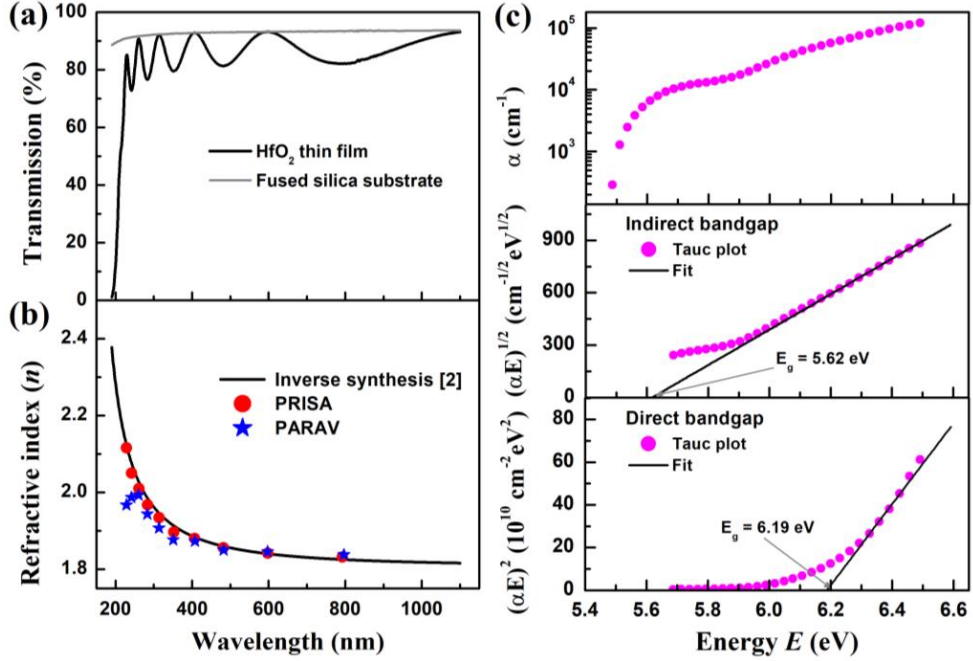

Fig. 5. (a) Transmission spectrum of the representative EB evaporated HfO$_2$ thin film sample H-3 along with that of fused silica substrate. (b) Refractive index of sample H-3 determined using PRISA, and PARAV are plotted along with the data obtained from reference [2]. (c) Absorption co-efficient (α) spectrum of sample H-3, and corresponding Tauc plots to determine indirect band gap ($E_g$ = 2.24 eV) and direct bandgap ($E_g$ = 2.41 eV) of the film using PRISA.

**Table-2: Parameters of EB evaporated HfO$_2$ thin films**

| HfO$_2$ thin film | O$_2$ pressure (mbar) | Thickness $d$ (nm) | | | $E_d$ (eV) | $E_0$ (eV) | $n_0$ | $E_g$ (eV) | |
|---|---|---|---|---|---|---|---|---|---|
| | | Fitting [2] | PRISA | PARAV | | | | Direct | Indirect |
| H-1 | 1x10$^{-4}$ | 260 | 264.4 ± 2.3 | 298 | 22.68 | 8.961 | 1.879 | 6.18 | 5.60 |
| H-2 | 2x10$^{-4}$ | 268 | 271.3 ± 3.2 | 295 | 22.13 | 8.94 | 1.864 | 6.18 | 5.61 |
| H-3 | 4x10$^{-4}$ | 322 | 324.7 ± 1.0 | 345 | 21.27 | 9.387 | 1.807 | 6.19 | 5.62 |
| H-4 | 8x10$^{-4}$ | 350 | 353.5 ± 3.5 | 369 | 20.12 | 9.481 | 1.767 | 6.21 | 5.64 |

Subsequently, dispersion energy parameters of the films are derived from the refractive index spectra using PRISA are listed in Table-2. The value of $E_d$ decreases from 22.68 eV to 20.12 eV, the values $E_0$ increases from 8.961 eV to 9.481 eV, and the value of $n_0$ decreases with increasing O$_2$ pressure. The variation of $n_0$ with O$_2$ pressure indicates that porosity in HfO$_2$ thin film increases with increasing O$_2$ pressure during deposition, as refractive index is proportional to the film density [49]. Algorithm in PRISA for estimating absorption co-efficient (α) of thin films at the interference-free absorption region uses equation (13). Subsequently, direct and indirect band gap of the films have been determined from the values of $α ≥ 10^4$ cm$^{-1}$ by fitting the Tauc plots using equation (14) for $M$ =1/2 and 2, respectively. The data of α vs.

$E$, $(\alpha E)^{1/2}$ vs. $E$, and $(\alpha E)^2$ vs. $E$, for the representative sample H-3 are plotted in Fig. 5 (c), are obtained using PRISA. The derived value of $E_g$ is 5.62 eV for indirect and 6.19 eV for direct band gap of the HfO$_2$ thin film (sample H-3). The bandgap $E_g$ of all the HfO$_2$ thin films are listed in Table-2. The value of $E_g$ increases with increasing O$_2$ pressure, and has the same trend like $E_0$, which validates the proportional correlation between them. The variation of $n_0$ and $E_g$ with O$_2$ pressure can be correlated via Moss rule: $n_0^4 E_g$=constant [50].

*4.3. Limitations*

PRISA gives accurate value of $n$, $k$, $d$, $E_d$, $E_0$, $n_0$, and $E_g$, for any kind of homogeneous thin films deposited on less dispersive transparent substrate and must have at least three interference maxima in its transmission spectrum. Absorption in the substrate affects the transmission spectrum of the thin film/substrate system, consequently PRISA gives inaccurate results (see supplementary material Fig. S2 and S3). However, the accuracy of the parameters can be significantly improved by analysing only the interference fringes data in the spectrum where the substrate seems to be transparent (see supplementary material Fig. S4). Refractive index inhomogeneity across the layer thickness influences the overall transmission spectrum of thin films (see supplementary material Fig. S5). It is found that PRISA provides reliable results of $n$, $k$, and $d$ for thin films having very small inhomogeneity (see supplementary material Fig. S6), while it fails for significantly inhomogeneous thin films (see supplementary material Fig. S7). Other than metallic thin films and very low thickness fringe free thin films, substrate absorption and film inhomogeneity are the other two factors that limit the use of PRISA for optical analysis of thin films.

## 5. Software distribution and future plans

PRISA 1.0 software is freely available and can be downloaded with a user manual from our website https://www.shuvendujena.tk/download. There is no requirement to install Python or any of its modules in order to run the software as it comes as a single executable (.exe) setup file, which can be directly run and installed on Windows-based computers. This setup file will not work on Linux/MAC operating system. We have planned to include analysis of reflection spectrum of thin films on absorbing or transparent substrates [12], and inverse synthesis of both reflection and transmission [51] in the next version of the software.

## 6. Conclusion

PRISA demonstrates reliable results for semiconductor and dielectric thin films. It determines $n$, $k$, $E_g$, $E_d$, $E_0$, and $d$ of thin films from their transmission spectra only. The thickness values of thin films obtained from PRISA are very much close to the true value within its error bar. It gives more accurate results as compared to PARAV software. PRISA is made simple and user-friendly aiming as a tool for researchers for rapid optical analysis of different thin film samples. The software can be freely downloaded from our website. The software can accurately analyse optical properties of any kind of thin films having spectra recorded above the fundamental absorption edge region with minimum three numbers of peaks in the transmission spectrum.

# Supplementary material

## PRISA: a simple software for determining refractive index, extinction co-efficient, dispersion energy, band gap, and thickness of semiconductor and dielectric thin films

S. Jena[*], R. B. Tokas, S. Thakur, D. V. Udupa

Atomic & Molecular Physics Division, Bhabha Atomic Research Centre, Mumbai 400 085, India

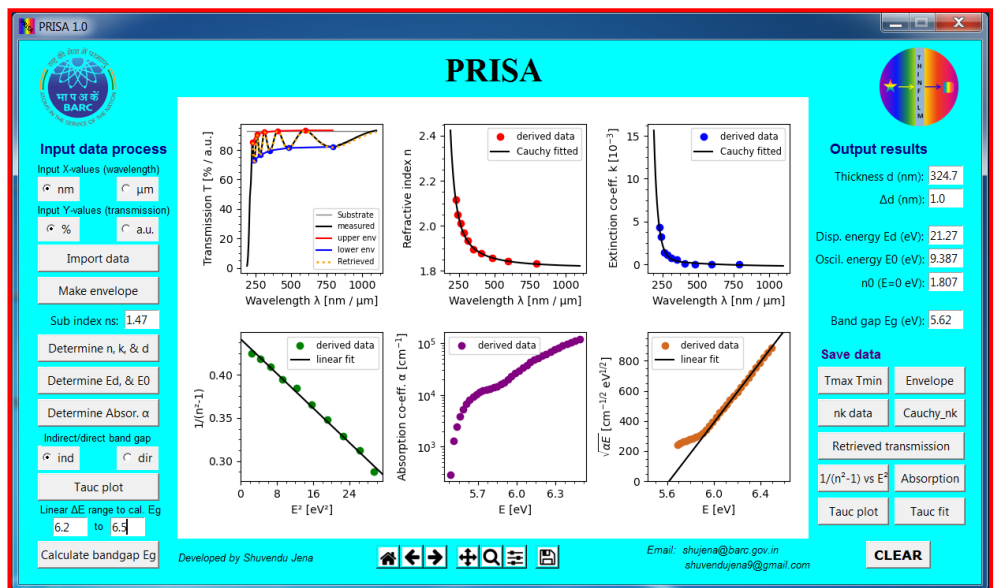

**Fig. S1.** GUI illustrating determination of $n$, $k$, $d$, $E_d$, $E_0$, and $E_g$ of electron beam evaporated $HfO_2$ thin films (sample H-3) [1] prepared at $O_2$ pressure of $4 \times 10^{-4}$ mbar.

## Absorption of substrate in NIR region:

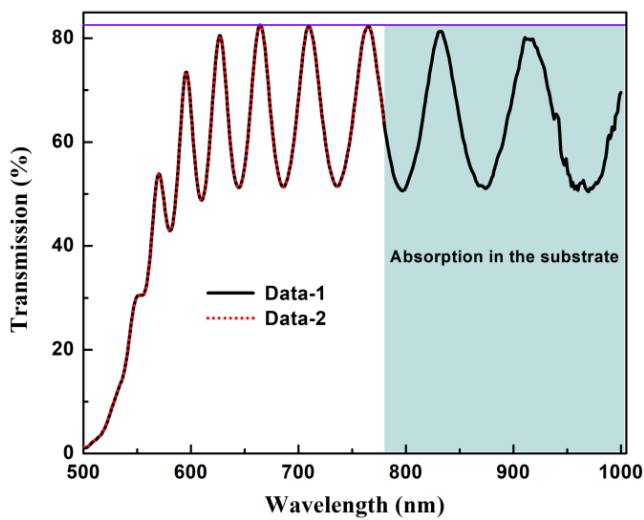

**Fig. S2.** Transmission spectrum of $Ge_{12}Sb_{25}S_{63}$ thin films [2]. Two types of data are extracted from the above spectrum such as Data-1 and Data-2. The possible absorption of the substrate in the NIR region is not considered in Data-2.

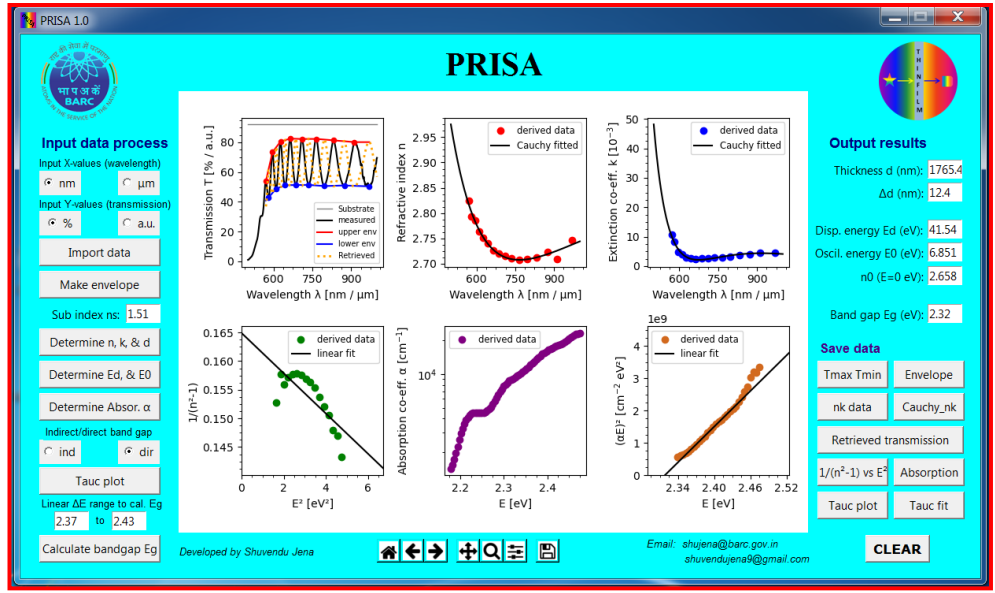

**Fig. S3.** GUI illustrating determination of $n$, $k$, $d$, $E_d$, $E_0$, and $E_g$ of laser irradiated $Ge_{12}Sb_{25}S_{63}$ chalcogenide thin film (Data-1). The retrieved transmission spectrum does not match with that of measured data which is due to wrong estimation of $n$, $k$, and $d$ of the film. Since the adopted Envelope method holds good for homogeneous thin films on transparent substrate, therefore the primary reasons for the mismatch could be either (i) the prepared thin film is inhomogeneous in nature, or (ii) the glass substrate has absorption in the NIR region. By closely analysing the data, it seems there exist absorption in the substrate, therefore we have removed the data in the absorption region and kept the data for only transparent region (Data-2) and then analyzed. We found that the retrieved spectrum now exactly overlapping the measured spectrum as shown in the figure below (Fig. S4).

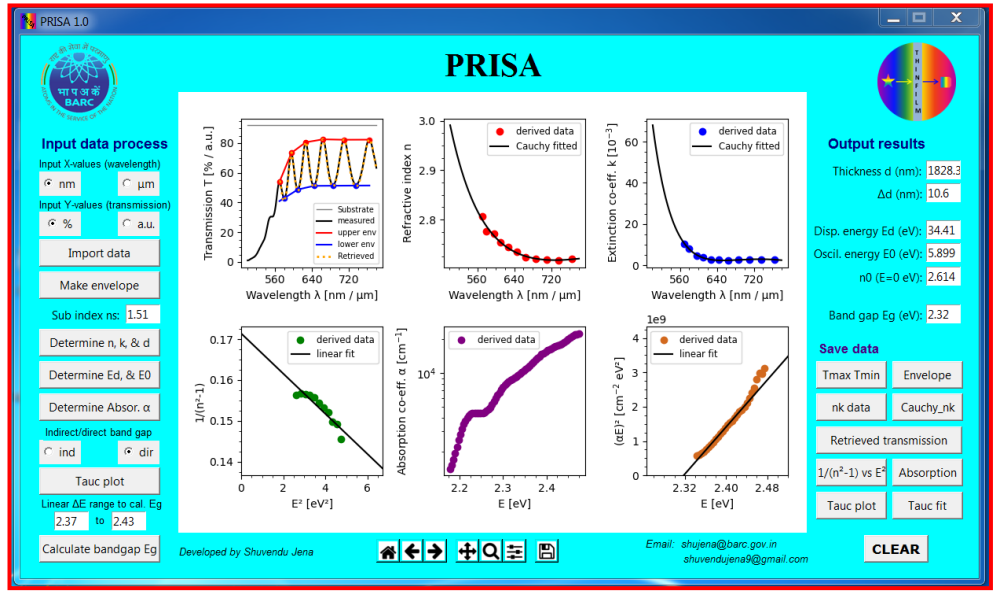

**Fig. S4.** GUI illustrating better analysis of $Ge_{12}Sb_{25}S_{63}$ thin films (Data-2), which provides more accurate value of $n$, $k$, $d$, $E_d$, $E_0$, and $E_g$, as the retrieved transmission exactly overlaps the measured spectrum of Data-2.

## Inhomogeneity in thin films:

Refractive index inhomogeneity or thickness inhomogeneity or both in a thin film affects its transmission and reflection spectrum. Examples of refractive index inhomogeneity in $HfO_2$ thin films are illustrated along with their schematics in the below figures. The optical transmission of the $HfO_2$ inhomogeneous thin films are numerically generated for two different inhomogeneity: (a) $\delta n = 0.3$ & $\delta d = 15$ nm, and (b) $\delta n = 0.3$ & $\delta d = 25$ nm. Refractive index (n) and extinction co-efficient (k) values of $HfO_2$ thin films used to numerically generate the spectra are taken from the reference [1], and thickness of bottom dense layer is taken as 400 nm, while the relatively top porous layer has very low thickness of 15 nm , and 25 nm, for the inhomogeneous structures shown in the inset of Fig. S5(a), and (b), respectively. Subsequently, these spectra are analyzed using PRISA to estimate the optical properties and thickness of the inhomogeneous thin films.

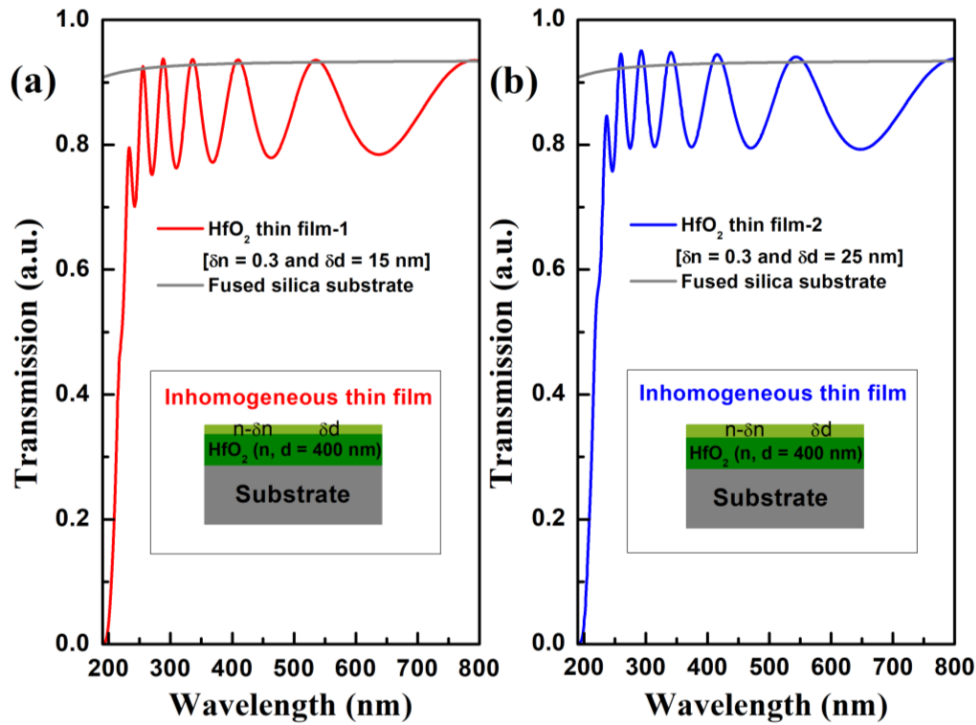

**Fig. S5.** Numerically generated transmission spectrum of inhomogeneous $HfO_2$ thin films for two different inhomogeneous structures: (a) $\delta n = 0.3$ & $\delta d = 15$ nm, and (b) $\delta n = 0.3$ & $\delta d = 25$ nm. The corresponding structures are drawn in the insets of the plot, and they are designated as $HfO_2$ thin film-1, and $HfO_2$ thin film-2, respectively.

Now the optical transmission spectrum of the two inhomogeneous thin films such as $HfO_2$ thin film-1 and $HfO_2$ thin film- 2 are analyzed using the software PRISA, and their complete GUI analysis are shown in Fig. S6, and S7, respectively. Analysis of $HfO_2$ thin film-1 reveals that PRISA can be used to analyse thin films having very small inhomogeneity and can provide reliable results. For thin films having significant inhomogeneity as in case of $HfO_2$ thin film-2, the software gives erroneous results for $n$, $k$, and $d$, subsequently, $E_d$, $E_0$, and $E_g$, which is illustrated in the Fig. S7.

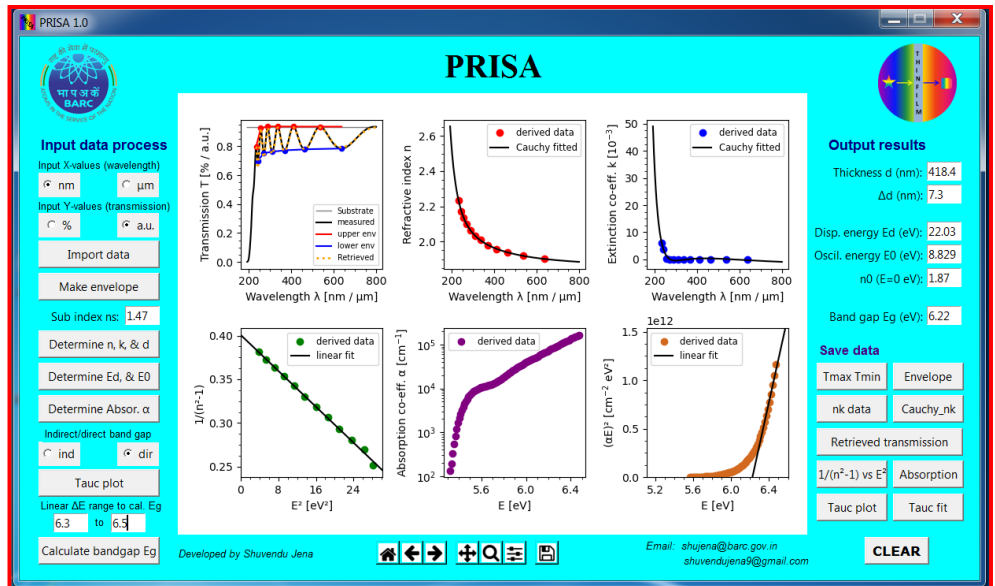

**Fig. S6.** GUI HfO$_2$ thin film-1 shows that the retrieved transmission exactly overlaps the original spectrum, therefore the obtained *n*, *k*, and *d* values can be considered reliable. The obtained thickness d = 418.4 ± 7.3 nm is almost matching to the original thickness of 415 nm.

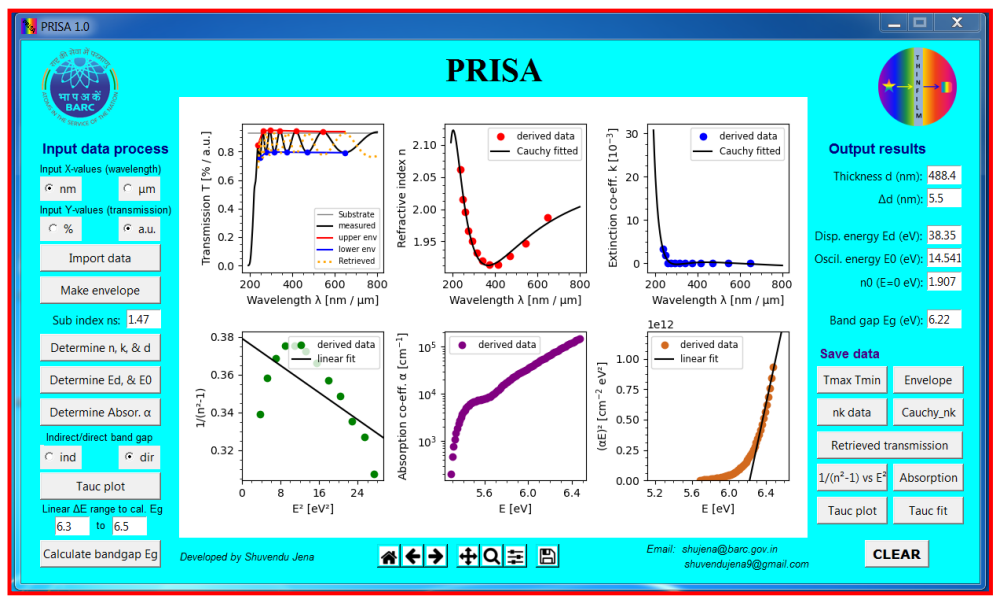

**Fig. S7.** GUI of HfO$_2$ thin film-2 shows that the retrieved transmission does not match with original spectrum, which is due to inaccurate value of *n*, *k*, and *d* obtained from the analysis. The obtained thickness d = 488.4 ± 5.5 nm is largely deviated from the original d = 425 nm.